\documentclass [preprint,
 amsmath,amssymb,
 aps,
 superscriptaddress,
12pt,floatfix]{revtex4-2}
\usepackage{graphicx} % Required for inserting images
\usepackage{makecell}
\usepackage{url} 
\usepackage{amsmath} % for math environments 
\usepackage{amssymb}
\usepackage{mathtools}
\usepackage{physics}
\usepackage{placeins}
\usepackage{enumitem}

\usepackage[T1]{fontenc} %to get underscore in Jessica's email
\usepackage{tikz}
\usetikzlibrary{calc}
\usetikzlibrary{arrows}
\usetikzlibrary{positioning}

\tikzstyle{susceptible} = [rectangle,  rounded corners, minimum width =1cm, minimum height =1cm,
text centered, text width=1cm, draw=black, fill=orange!35, line width = 0.6 mm]
\tikzstyle{process} = [rectangle,  rounded corners, minimum width =1cm, minimum height =1cm,
text centered, text width=1cm, draw=black, fill=blue!17, line width = 0.6 mm]
\tikzstyle{recovered} = [rectangle,  rounded corners, minimum width =1cm, minimum height =1cm,
text centered, text width=1cm, draw=black, fill=green!30, line width = 0.6 mm]
\tikzstyle{empty} = [node distance=3cm]	
\tikzstyle{arrow}= [thick, scale=1.3, ->, >=triangle 60]
\tikzstyle{arrow2}= [draw=gray!70, scale=1.3, ->, >=triangle 45]

\usepackage{lineno}
\usepackage{todonotes}
\definecolor{purple1}{rgb}{0.949, 0.941, 0.969}
\definecolor{purple5}{rgb}{0.329, 0.153, 0.561}
\definecolor{green1}{HTML}{edf8fb}
\definecolor{green5}{HTML}{006d2c}
\definecolor{pink1}{HTML}{f1eef6}
\definecolor{pink4}{HTML}{dd1c77}
\definecolor{blue1}{HTML}{D6EAF8}
\definecolor{blue2}{HTML}{3357FF}

\newcommand{\JCM}[1]{\todo[
    color=orange,
    bordercolor=black,
    size=\small,
    inline]{#1 --- Joel}} %for comments by Joel

\newcommand{\cz}[1]{\todo[
  color=green1,
  bordercolor=green5,
  size=\small,
  inline
  ]{\textcolor{green5}{CZ: #1 }}}

\newcommand{\jes}[1]{\todo[
  color=pink1,
  bordercolor=pink4,
  size=\small,
  inline
  ]{\textcolor{pink4}{JES: #1 }}}

\newcommand{\Ro}{\mathcal{R}_0} %R_0
\newcommand{\loss}{\mathcal{L}} 

\begin{document}

\title{Including frameworks of public health ethics in computational modelling of infectious disease interventions}

\author{Alexander E. Zarebski}
\thanks{These authors contributed equally}
\affiliation{School of Mathematics \& Statistics, University of Melbourne, Australia}
\affiliation{Pandemic Sciences Institute, University of Oxford, United Kingdom}
\affiliation{MRC Biostatistics Unit, University of Cambridge, United Kingdom}

\author{Nefel Tellioglu}
\thanks{These authors contributed equally}
\affiliation{Department of Infectious Diseases, University of Melbourne, at the Peter Doherty Institute for Infection and Immunity, Melbourne, Victoria, Australia}

\author{Jessica E. Stockdale}
\affiliation{Department of Mathematics, Simon Fraser University, Burnaby, Canada}

\author{Julie A. Spencer}
\affiliation{Information Systems and Modeling Group, Los Alamos National Laboratory, Los Alamos, New Mexico, United States}

\author{Wasiur R. {KhudaBukhsh}}
\affiliation{School of Mathematical Sciences, University of Nottingham, United Kingdom}

\author{Joel C. Miller}
\affiliation{Department of Mathematical and Physical Sciences, La Trobe University, Australia}
\affiliation{Australian Centre for AI in Medical Innovation}

\author{Cameron Zachreson}
\email{cameron.zachreson@unimelb.edu.au}
\affiliation{School of Computing and Information Systems, University of Melbourne, Parkville, Victoria, Australia}

\date{\today}

\begin{abstract}
Decisions on public health interventions to control infectious disease are often informed by computational models. Interpreting the predicted outcomes of a public health decision requires not only high-quality modelling, but also an ethical framework for assessing the benefits and harms associated with different options. The design and specification of ethical frameworks matured independently of computational modelling, so many values recognised as important for ethical decision-making are missing from computational models. We demonstrate a proof-of-concept approach to incorporate multiple public health values into the evaluation of a simple computational model for vaccination against a pathogen such as SARS-CoV-2. By examining a bounded space of alternative prioritisations of values (outcome equity and aggregate benefit) we identify value trade-offs, where the outcomes of optimal strategies differ depending on the ethical framework. This work demonstrates an approach to incorporating diverse values into decision criteria used to evaluate outcomes of models of infectious disease interventions. 
\end{abstract}

\maketitle

% \begin{table}[h!]
% \begin{tabular}{lllll}
% \hline
% Name & Email & ORCID & Affiliations & Acknowledgement \\ \hline
% x    & x     & x     & x            & x               \\
% Nefel Tellioglu    & tellioglu.n@unimelb.edu.au     & 0000-0002-6719-5960     & e            & x               \\
% Joel C. Miller & joel.miller.research@gmail.com & 0000-0003-4426-0405 & a,i& x\\
% Alexander E. Zarebski  & azarebski@unimelb.edu.au     &  0000-0003-1824-7653    & b,c,j            & x               \\
% Cameron Zachreson    & cameron.zachreson@unimelb.edu.au     & 0000-0002-0578-4049     & d            & x               \\
% Julie A. Spencer & jaspencer@lanl.gov & 0000-0001-8756-6987 & g & x \\
% Jessica E. Stockdale & jessica\_stockdale@sfu.ca & 0000-0001-7984-1010 & f  & x\\
% Wasiur R. KhudaBukhsh & wasiur.khudabukhsh@nottingham.ac.uk & 0000-0003-1803-0470 & h & x\\ \hline
% \end{tabular}
% \end{table}

\newpage

\section{Introduction}

% \aez{I think another citation earlier in this paragraph that points to a ``public health ethics for STEM dummies'' would be a useful addition. I am not familiar with the literature, perhaps a chapter from \emph{Philosophical Perspectives on Bioethics}? Something that allows a computer nerd to have a sensible starting point to encounter the fundamental definitions/ideas. Again, I think it would be really nice to spell out the difference between equality and equity in a technical sense. I don't have the words to do a good job of this though.}

Frameworks of public health ethics offer guidelines for balancing competing values when making important decisions that affect the lives and livelihoods of populations \cite{mastroianni2019oxford,upshur2019oxhand}. In responding to infectious disease outbreaks and pandemics, public health interventions have often prioritised preventing clinical burden from overwhelming health services. This corresponds to utilitarian ethical frameworks seeking to maximise aggregate well-being, without regard for the distribution of goods and harms \cite{savulescu2020utilitarianism}.

%\aez{I think it would be \textbf{very} useful to describe the difference between \emph{equality}, i.e. distributing vaccines uniformly, and \emph{equity}, i.e. distributing vaccines so as to have uniformly distributed consequences. I would like this contrast to appear in the Discussion, but I think it is more important to get it in early in the Introduction}

%\jas{Alex, here's an attempt at implementing your suggestion. Please edit as needed.}
%\cz{this is nice but it should be moved to come after the discussion of utilitariansim - it's a more subtle point}
%Maximizing aggregate well-being requires careful definition. Recent work in the field of ethics makes it clear that equality and equity of interventions can have disparate consequences \cite{curran2022must}. For example, distribution of vaccines uniformly to the maximum number of people (equality), while a common public health goal, may result in an uneven distribution of outcomes (inequity), which can depend on how well-being is understood \cite{giubilini2021queue}.

Evaluating public health decisions by aggregate benefit has the potential to create or exacerbate inequitable distributions of benefits and burdens \cite{rao2020should,lancet2020india,Haumaru2021report}. To address this issue in utilitarian public health policy, governments and public health institutions have made efforts to design ethical frameworks to balance aggregate benefits and equitable outcomes \cite{osullivan2022ethical,world2007ethical,fielding2021constructing, malmusi2022multi, boden2017model,upshur2019oxhand}. 

During the COVID-19 pandemic, computational and mathematical models were widely applied to inform public health interventions. While modelling provided invaluable guidance to policy makers, it typically did not account for impacts other than aggregate clinical burden, ignoring other values relevant to public health ethics, such as equity \cite{zachreson2024ethical}. Decision makers had to make judgments on these impacts without quantitative guidance. To help address this deficiency, modellers need to translate qualitative features of ethical frameworks into quantitative model features. 

Here, we demonstrate a way to include some elements of ethical frameworks within a quantitative model of vaccination against a pandemic pathogen.
We consider a simple model of pathogen transmission to illustrate encoding ethical values into an optimisation problem for selecting a vaccination strategy.
Including these ethical frameworks into the optimisation problem allows us to balance (potentially) competing values. Here, we choose to examine outcome equity and aggregate clinical burden, as illustrative examples. To connect our approach to a realistic case study, we implement a model of COVID-19 transmission using parameters from the literature.
% Our choice of ethical values and pathogen characteristics requires the model to have two properties, which we expect to be present in most scenarios:
% \vspace{-7mm}
% \begin{enumerate}
% \item The population is heterogeneous (in our example, there are two age groups with different risk profiles);\vspace{-7mm}
% \item The intervention may cause harm (in our example, vaccination has limited efficacy and can produce clinical side effects in rare instances.)
% \end{enumerate}
% \vspace{-7mm}
% \noindent
% The first assumption is necessary because it allows us to quantify outcome equity across two subpopulations. 
% The second assumption is necessary because with a perfect intervention, using the intervention is trivially the optimal solution.
% These assumptions add complexity, but we are confident they are nearly universally relevant. 

Below, in section~\ref{sec:methods}, we describe our model of pathogen transmission with a variable level of vaccination in an age-stratified population. 
We cast the selection of a vaccination strategy as an optimization problem, selecting a level of vaccination that minimises the value of a loss function which provides a quantitative description of how ``good'' a given intervention is.
We demonstrate an approach to encoding qualitative ethical frameworks into the quantitative loss function. This connects values derived from ethical frameworks in public health to model design choices, as suggested by Zachreson et al. \cite{zachreson2024ethical}.
While the loss function we describe is specific to this model and choice of ethical values, the approach may be generalised.

In section~\ref{sec:results}, we describe the parameterization of our model that we use to illustrate this approach for a scenario based on COVID-19 in an urban population (Melbourne, Australia).
We first analyse the optimal level of vaccination in each age group under three (extreme) ethical frameworks, and then expand our analysis to a broader range of prioritisation schemes for ethical values. We perform this analysis for two scenarios: one with unlimited vaccine supply and another with supply limited to 20\% of the total population. For each scenario, we characterise the properties of optimal interventions over the space of ethical frameworks we consider. We then describe the qualitative and quantitative features of our results, including the presence or absence of trade-offs between values. We conclude with a discussion of the key findings, and the limitations of our study. 
%For ethical frameworks prioritising the reduction of aggregate clinical burden (i.e. close to utilitarianism), there is a robust optimal strategy: vaccinate as many people as you can, prioritising the elderly.

\section{Methods}\label{sec:methods}

\subsection{Transmission model}

To simulate pathogen transmission and mitigation, we use a deterministic SIR model with imperfect vaccination (using the ``all-or-none'' mode of vaccine failure).
We consider a heterogeneous population consisting of $N$ people divided into two groups of size $N_1$ and $N_2$.
Contact occurs within and between groups (at potentially different rates).
Each group can be further divided into those who are susceptible to infection, currently infectious, or removed from the infectious class (and hence immune to subsequent infection). 

Let $S_i(t)$, $I_i(t)$, and $R_i(t)$ denote the number of people in group $i$ who are susceptible, infectious, and immune at time $t$.
Initially, there is a single infected individual (index case) in each group. 
Members of $S_i$ become infected and move to $I_i$ at a rate $\sum_j \beta_{ji} I_{j}/N_{i}$. 
The $j$th summand corresponds to an infected individual of group $j$ infecting an individual of group $i$.
Infection occurs at rate $\beta_{ji}=\beta c_{ji}$ where $\beta$ is a global transmission scalar and $c_{ij}$ is the per-capita contact rate between members of group $i$ and $j$.
Infection occurs if the individual in group $i$ is susceptible, which has probability $S_i/N_i$.
Members of $I_i$ recover from infection (and become immune to subsequent infection) at rate $\gamma$.
Although we include only two groups ($i, j \in \{1, 2\}$), the model can be readily extended to incorporate additional heterogeneity.

\subsubsection{Imperfect vaccination}

We assume an ``all-or-none'' mode of vaccine failure in which a proportion of people receiving the vaccine gain full immunity, and the remainder are not protected against infection.
For simplicity, we assume all vaccinations occur before the initiation of the outbreak, with proportions $p_1$ and $p_2$ of individuals in groups 1 and 2 being vaccinated.
Each choice of how many people to vaccinate in each group, $p_i$, is a \emph{vaccination strategy}. The selection of a vaccination strategy is the focus of our optimisation framework (see below). 
 
\begin{figure}
    \centering
    \includegraphics[width=0.95\linewidth]{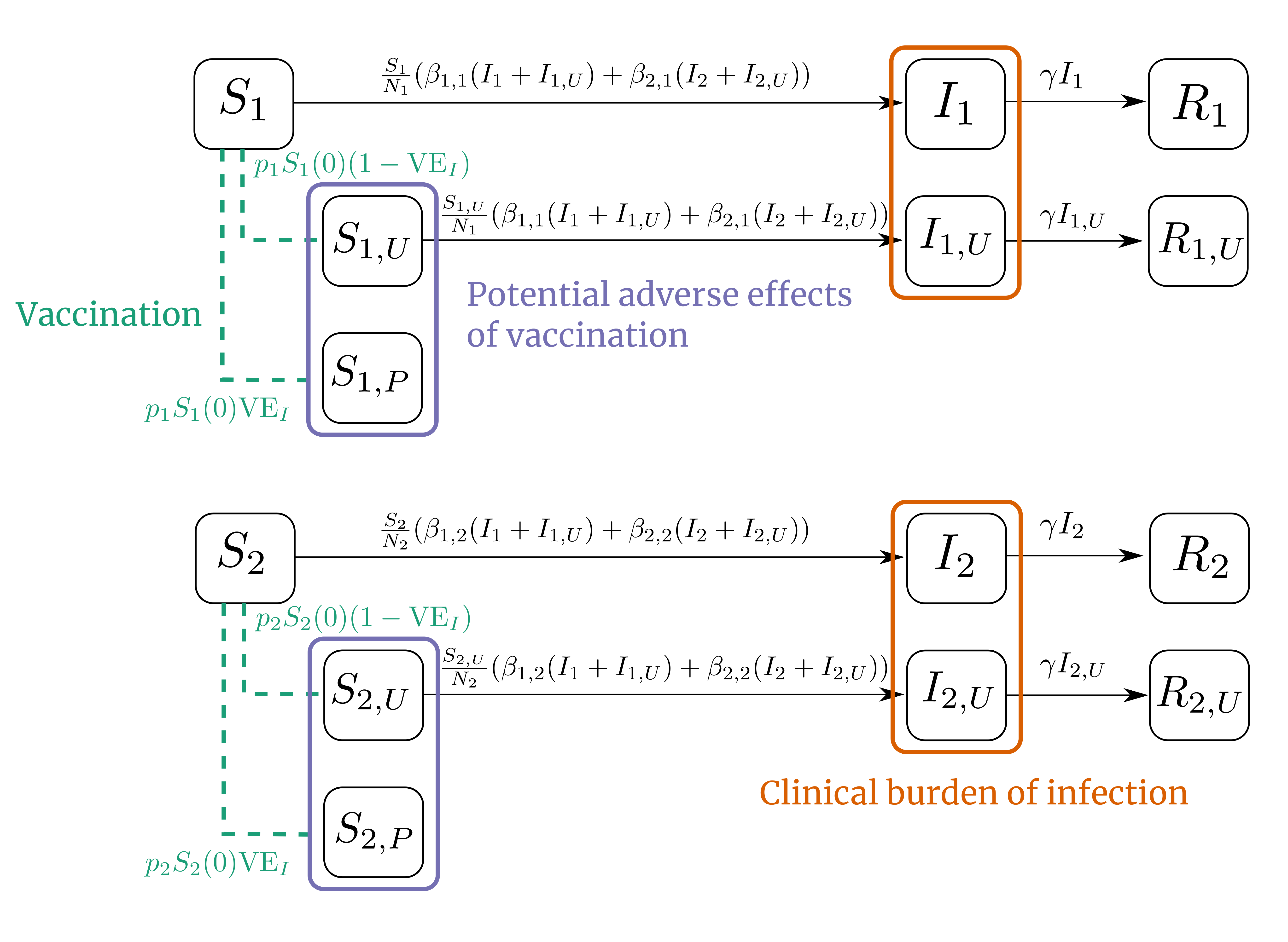}
    \internallinenumbers
    \caption{Compartmental model diagram for the two subpopulations with imperfect ``all-or-none'' vaccination indicating which compartments contribute to the  components of the loss function.
    Each group is stratified into susceptible $S$, infectious $I$ and removed $R$ (i.e. immune due to recovery.)
    Subscripts represent group membership and vaccination status: vaccinated but unprotected $U$, or vaccinated and protected $P$.
   Dashed green lines indicate the proportion initially vaccinated. The purple and orange boxes indicate the compartments contributing to adverse effects of vaccination and clinical burden.
   }
    \label{fig:ODE_vacc}
\end{figure}

In our transmission model, unvaccinated individuals are mathematically the same as those who are vaccinated but unprotected (through vaccine failure). Both are equally susceptible to infection and, once infected, equally infectious. That is, we assume no partial protection against infection or onward transmission.
We use subscripts to indicate vaccination status.
For example, in group $i$, the members of $S_{i}$ are unvaccinated, $S_{i,U}$ are vaccinated but \emph{unprotected}, and $S_{i,P}$ are vaccinated and \emph{protected}. The compartmental transmission model with stratification by group and vaccination status is shown in Figure~\ref{fig:ODE_vacc}.
The full system of ODEs corresponding to the diagram in Figure~\ref{fig:ODE_vacc} is provided in the Supporting Information Equation \ref{eq:ode_model_with_vacc}.

\subsection{Components of the loss function}

To choose the best vaccination strategy (choice of $p_1$ and $p_2$)  under a given ethical framework, it is necessary to measure the {\it{loss}} associated with the outcome.
To do this, we extend the transmission model described above by adding \emph{clinical burden}.
We distinguish between two types of clinical burden: the clinical burden due to severe disease caused by the pathogen, (which is reduced by the vaccine-induced protection even when vaccines fail to protect against infection); and the clinical burden due to rare adverse side-effects of vaccination, as described below.

% The task of selecting an intervention in this model corresponds to selecting the proportions $p_1$ and $p_2$, which specify the number of individuals vaccinated in groups 1 and 2, respectively.
% Once the intervention parameters are selected, we can simulate the corresponding scenario using the model described above. 
% In practice, one seeks a choice of $p_1$ and $p_2$ that results in the most favorable outcome. Here, we are interested in comparing the outcomes of optimal strategies selected under different ways of prioritising values, which we refer to as ethical frameworks.

The modelling literature tends to focus on selecting interventions that minimise the aggregate clinical burden.
However, there is a growing recognition that this should also consider the equity of clinical burden across the population.
In particular, where groups include populations who already face disproportionate burdens in other domains.
As such, we consider two measures of equity across the two populations: equity of burden from infection outcomes, and equity of burden from the adverse impacts of vaccination.
Below we describe a family of loss functions. Each loss function encodes an {\it{ethical framework}}, which we define for the purposes of this study as a precise prioritisation of the values we include (see below for details). 

\subsubsection{Loss function components: clinical burden}

The clinical burden due to infection in group $i$ (e.g. the total number of days members of group $i$ spend in hospital) is $\loss^{I}_{i}$.
This accounts for the infections among both for unvaccinated individuals and those who are vaccinated but unprotected:
\begin{equation}
    \loss^I_i = C^I_i R_i(\infty) + C^I_{i,U}R_{i,U}(\infty) \,,
\end{equation}
\noindent
where $R_{i}(\infty)$ and $R_{i,U}(\infty)$ are the numbers of unvaccinated and vaccinated-unprotected individuals infected in group $i$ across the whole epidemic and the $C^I_i$ and $C^I_{i,U}$ are the average clinical burden for unvaccinated and vaccinated-unprotected members of group $i$.
We allow for the vaccine to reduce the severity of breakthrough infections by specifying $C^I_{i, U} = C^I_i(1 - \text{VE}_{SD|I})$.
The term $\text{VE}_{SD|I}$ is a scaling factor for the vaccine's efficacy against severe disease, given infection.
The aggregate clinical burden due to infection in both groups is:
\begin{equation}
\loss^I = \loss^I_1 + \loss^I_2\,.
\end{equation}

To include the clinical burden of adverse effects of vaccine in our decision framework, we compute the expected clinical burden due to vaccination in group $i$ as
\begin{equation}
\loss^V_i = C^V_i V_i\,,
\end{equation}
where $C^V_i$ is the average clinical burden (e.g. expected days in hospital) due to adverse effects of vaccine for a member of group $i$, and $V_i = N_ip_i$ is the total number of vaccinations administered to members of group $i$.
The net clinical burden associated with vaccination is:
\begin{equation}
\loss^V = \loss^V_1 + \loss^V_2\,.
\end{equation}

Combining the contributions of infection-induced and intervention-induced clinical burden we get the aggregate clinical burden \(\loss_{CB}\) which is given by
\begin{equation}\label{eq:lcb}
\loss_{\text{CB}} = \loss^I + \loss^V\,.
\end{equation}

We note that this aggregate clinical burden that is a typical target of optimisation when selecting vaccination strategies.
To account for considerations of equity, we need to consider another component to the loss function as described below.

\subsubsection{Loss function components: equity}

Here, we consider an equitable vaccination strategy to be one in which the per-capita clinical burden is the same for both groups. This is in contrast to prioritising equity of vaccine distribution, which would see the vaccine supply spread uniformly across all members of the population.

We start by considering what outcome would represent equality of per-capita burden across groups.
If the population shared clinical burden due to infection uniformly the loss experienced by group \(i\) would be \(\loss^IN_i/N\), since \(N_i/N\) is the proportion of the population in group \(i\).
Similarly, if the burden of adverse effects of vaccination were uniformly distributed, the loss experienced by group \(i\) would be \(\loss^V(N_i/N)\).

To quantify how equitable an intervention is, we measure how close the outcome is to equality. We compute the following loss function measuring the equality of infection burden per-capita:
% \begin{equation}
% \loss_{\text{EI}}= \left \lVert (\loss^I_1,\loss^I_2) - \loss^I(N_1,N_2)/N \right \rVert_{1}
% \end{equation}
\begin{equation}\label{eq:lei}
\loss_{\text{EI}}= \left \lVert \bigg{(}\loss^I_1 - \loss^I\frac{N_1}{N}~, ~\loss^I_2 - \loss^I\frac{N_2}{N}\bigg{)} \right \rVert_{1}
\end{equation}

\noindent
where \(\left \lVert \cdot \right \rVert_{1}\) is the \(L^{1}\)-norm, that is \(\left \lVert (x_1,x_2) \right \rVert_{1}=\abs{x_1}+\abs{x_2}\).
The value of \(\loss_{\text{EI}}\) measures how far the observed distribution of burden \((\loss^I_1,\loss^I_2)\) is from an equal distribution of burden \((\loss^I\frac{N_1}{N},\loss^I\frac{N_2}{N})\). Similarly, the measure of equality in the distribution of adverse effects of vaccination is given by
\begin{equation}\label{eq:lev}
\loss_{\text{EV}}= \left \lVert \bigg(\loss^V_1 - \loss^V\frac{N_1}{N}~, ~\loss^V_2 - \loss^V\frac{N_2}{N}\bigg) \right \rVert_{1}.
\end{equation}

The terms \(\loss_{\text{EI}}\) (Equation \ref{eq:lei}) and \(\loss_{\text{EV}}\) (Equation \ref{eq:lev}) measure how far away the per-capita outcomes are from equality for a given combination of $p_1$ and $p_2$.
In the next section we describe our approach to combining these terms with the aggregate clinical burden, \(\loss_{\text{CB}}\) (Equation \ref{eq:lcb}), for a single loss function.

\subsubsection{Combined loss function}

Note that \(\loss_{\text{CB}}\), \(\loss_{\text{EI}}\) and \(\loss_{\text{EV}}\) take values over different units and ranges.
We normalise each of these terms to obtain dimensionless quantities taking values from \(0\) to \(1\).
Since we are optimising over a bounded parameter space ($p_1$ and $p_2$ take values between zero and one) we can numerically compute the range of values for each loss term. This range of values can then be used to normalise the term to take values between zero and one: 
\begin{equation}
  \loss_{X}^*=\frac{\loss_{X}-\min(\loss_{X})}{\max(\loss_{X}) - \min(\loss_{X})}\,,  
\end{equation}
where $X$ represents a loss-term (subscript CB, EI, and EV), and the asterisk (*) indicates the value has been normalised to the range [0, 1].

Our loss function is a linear combination of the three normalised loss terms with a weighting of the terms specified by $w_{\text{EI}}$ and $w_{\text{EV}}$:
\begin{equation}\label{eq:GLoss}
\loss = (1 - w_{\text{EI}} - w_{\text{EV}})\loss_{\text{CB}}^{*} + w_{\text{EV}} \loss_{EV}^{*} + w_{\text{EI}} \loss_{\text{EI}}^{*}.
\end{equation}
The value of $w_{\text{EV}}$ corresponds to priority given to equity in vaccine-induced adverse outcomes, $w_{\text{EI}}$ corresponds to the priority given to equity of infection-induced clinical outcomes, and $(1 - w_{\text{EI}} - w_{\text{EV}})$ is the priority given to aggregate clinical burden due to infection. Note that we require the weights to satisfy the following properties: \(0\leq w_{\text{EI}} \leq 1\,, \text{and}~0\leq w_{\text{EV}} \leq 1\) and \(w_{\text{EI}} + w_{\text{EV}}\leq 1\).

We can now define an \emph{ethical framework} quantitatively as a choice of $w_{\text{EV}}$ and $w_{\text{EI}}$ which reflects our desired prioritisation of these factors.
For example, a utilitarian perspective would be realised by small values of $w_{\text{EI}}$ and $w_{\text{EV}}$ (i.e., $1 - w_{\text{EI}} - w_{\text{EV}} > w_{\text{EI}} + w_{\text{EV}}$).
This would result in  the optimisation of vaccination primarily minimising aggregate clinical burden.
From the perspective of someone interested in equitable outcomes, one might select larger values for $w_{\text{EI}}$, producing a greater penalty for inequality in the infection-induced burden.
Further, a policy maker may have good reason to prioritise equity in terms of harms associated with the intervention (induced harms) over equity in harms associated with infections.
Because these harms arise as a directly attributable result of the intervention, they may be more predictable and more easily ascribed to the policy choice than harms associated with infections.
% In such a case, a policy maker could choose to constrain their choices such that $w_{\text{EV}} > w_{\text{EI}}$. 
The space of possible ethical frameworks defined by the family of loss functions in Equation \ref{eq:GLoss} is depicted by the diagram in Figure~\ref{fig:schematic_Obj_space}.

\subsection{Example: COVID-19}

To investigate the application of our approach under a realistic set of parameters, we compare the results of applying different ethical frameworks to vaccination against COVID-19.
Our model is parameterised to match the population of Melbourne, Australia, during the Omicron phase of the pandemic from late 2021 through mid-2022. 
Motivated by the increased severity of COVID-19 for older adults, including longer average hospital stays \cite{tobin2023real}, we divide our population into groups corresponding to those aged 0-69 years and those aged 70 years or older.
We consider two different scenarios: one in which the supply of vaccines is unlimited; and another with a limited number of vaccines, as might occur early in an outbreak of a novel pathogen before the logistics of production and distribution have been established. We note that while our parameter estimates are broadly realistic, it is not our intention to exactly reproduce the dynamics of COVID-19 in Australia, which would require a much more detailed model. 

\subsubsection{Parameterisation}

Parameters used for our model of COVID-19 in Australia during the Omicron period are listed in Table \ref{tab:sir2-params}.
These values are derived from census data and parameter estimates from the epidemiological literature which are summarised in the Supporting Material, Tables ~\ref{tab:params1}, \ref{tab:params2}, and \ref{tab:params3}.
The main features of this parameterisation are:
\begin{itemize}[itemsep=0cm]
\item{a relatively high reproductive ratio (\(\Ro=3.4\)),}
\item{a modest vaccine efficacy against infection,}
\item{a higher (but still modest) vaccine efficacy against severe disease given infection,}
\item{a much higher risk of severe infections in group 2 (representing people 70 years of age or older) than group 1,}
\item{and a higher risk of adverse impacts of vaccination in group 1 than group 2 (driven by a higher risk of myocarditis in men aged under 30);}
\end{itemize}

\begin{table}[]
\centering
\begin{tabular}{l | l | l}
\hline
Parameter                          & Group 1 value   & Group 2 value  \\
\hline
Population size, $N_i$             & 4,395,000       & 605,000        \\
%Age range                          & $0$--$69$       & $70+$          \\

$\Ro$                              & 3.4            & 3.4             \\

$\beta$                            & 0.564          & 0.564           \\

contact matrix            & \makecell[l]{$c_{11} = 0.38$,\\ $c_{12} = 0.14$} & \makecell[l]{$c_{22} = 0.34$,\\ $c_{21} = 0.14$} \\

$\gamma$                           & 0.096          & 0.096           \\

Initially infected, $I_i(0)$       & 1               & 1              \\ 

$\text{VE}_I$                      & 0.531           & 0.531          \\

$\text{VE}_{SD|I}$                   & 0.627        & 0.627           \\

Expected infection cost, $C^I_i$   & $0.00088\times 2.87$~~& $0.0104\times 7.61$~~\\

Expected vaccination cost, $C^V_i$~ & $6\times 10^{-5}\times 5.7$ & $2\times 10^{-5}\times 5.7$    \\ \hline

\end{tabular}
\internallinenumbers
\caption{Parameters of the transmission and clinical burden models for COVID-19 based on the estimates from Tables~\ref{tab:params1}, \ref{tab:params2} and \ref{tab:params3}. See the Supporting information for more details on parameter estimates.} 
\label{tab:sir2-params} 
\end{table}

\subsubsection{Computational method for optimising the vaccination strategy}

To locate optimal vaccination strategies for the model and loss functions described above, we perform a grid-search of $\loss$ over $p_1$ and $p_2$ with a step size of 0.02 for scenarios with unlimited vaccine supply and a step size of 0.01 for scenarios with limited vaccine supply (see below).

Due to discretisation effects, the grid search algorithm does not guarantee that the global minimum will be located (this is visible in Figure~\ref{fig:traj_loss_unlimited}(f) below). However, plots of the full loss surface suggest that the true optimum is near enough to the grid search result that our findings are not impacted substantially by discretisation effects.

We note that in the case where \(w_{\text{EV}}=1\), a unique optimal vaccination strategy does not exist because all equal per-capita distributions of adverse impacts produce equivalent results (Figures \ref{fig:traj_loss_unlimited}f and \ref{fig:traj_loss_limited}f illustrate this behaviour). As such, we have excluded the corresponding point from the space of ethical frameworks that we consider below (Figure~\ref{fig:hmaps_unlimited} and Figure~\ref{fig:hmaps_limited}).

\subsubsection{Ethical frameworks}

\begin{figure}
    \centering
    \includegraphics[width=0.6\linewidth]{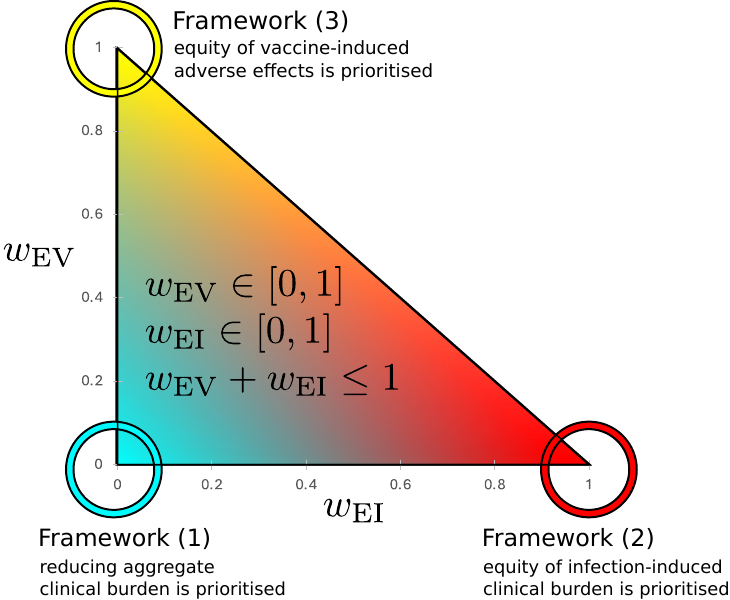}
    \internallinenumbers
    \caption{The space of loss functions used to define different ethical frameworks. The two weight coefficients $w_{\text{EI}}$ and $w_{\text{EV}}$ determine the prioritisation of three values we consider: 1) aggregate clinical burden, 2) equity of infection-induced clinical burden, and 3) equity of clinical burden from vaccine-induced adverse reactions. The three extreme ethical frameworks we use as examples in our results are indicated by the teal, red, and yellow circles at the corners of the triangle.}
    \label{fig:schematic_Obj_space}
\end{figure}

We consider the optimal vaccination strategy (choice of $p_1$ and $p_2$) under three specific ethical frameworks, with each representing one of three extremes:
\begin{enumerate}
\item{Framework 1 ($w_{\text{EI}} = 0, ~w_{\text{EV}} = 0$): exclusively prioritises total clinical burden (Figures \ref{fig:traj_loss_unlimited}a and \ref{fig:traj_loss_unlimited}d).}
\item{Framework 2 ($w_{\text{EI}} = 0.99, ~w_{\text{EV}} = 0$): prioritises equity of infection-associated clinical burden almost exclusively, with a small weight fraction $1\%$, given to aggregate clinical burden (Figures \ref{fig:traj_loss_unlimited}b and \ref{fig:traj_loss_unlimited}e).}
\item{Framework 3 ($w_{\text{EI}} = 0, ~w_{\text{EV}} = 0.99$): prioritises equity of clinical burden associated with adverse impacts of vaccination almost exclusively, with a small weight fraction $1\%$, given to aggregate clinical burden (Figures \ref{fig:traj_loss_unlimited}c and \ref{fig:traj_loss_unlimited}f).}
\end{enumerate}
The positioning of these three frameworks, relative to the overall space of potential loss functions, is shown in Figure~\ref{fig:schematic_Obj_space}.
In frameworks two and three, we set $w_{\text{EI}}$ and $w_{\text{EV}}$ (respectively) close but not equal to one.
This ensures the optimisation algorithm chooses strategies that produce equitable outcomes while ensuring that if there are multiple equivalently equitable scenarios, the optimization chooses the one corresponding to the lowest clinical burden.
We note that these frameworks 2 and 3 are extreme, in the sense that they put little weight on minimising aggregate clinical burden, but they useful in illustrating the types of trade-off that may emerge.

\FloatBarrier

\section{Results}\label{sec:results}

After describing the results of optimising interventions subject to loss functions corresponding to frameworks 1--3, we examine intermediate weightings over a broad range of values for $w_{\text{EI}}$ and $w_{\text{EV}}$.
This allows us to characterise the decision space (Figure~\ref{fig:schematic_Obj_space}) in terms of the chosen vaccination strategies and the associated outcomes. By doing so, we identify potential trade-offs between equity and aggregate clinical burden. Below, we provide this analysis for two scenarios, one in which vaccine supply is unlimited, and one for which vaccine supply is limited to only 20\% of the total population. 

\subsection{Unlimited vaccine supply}

\begin{figure}
    \centering
    \includegraphics[width=0.95\linewidth]{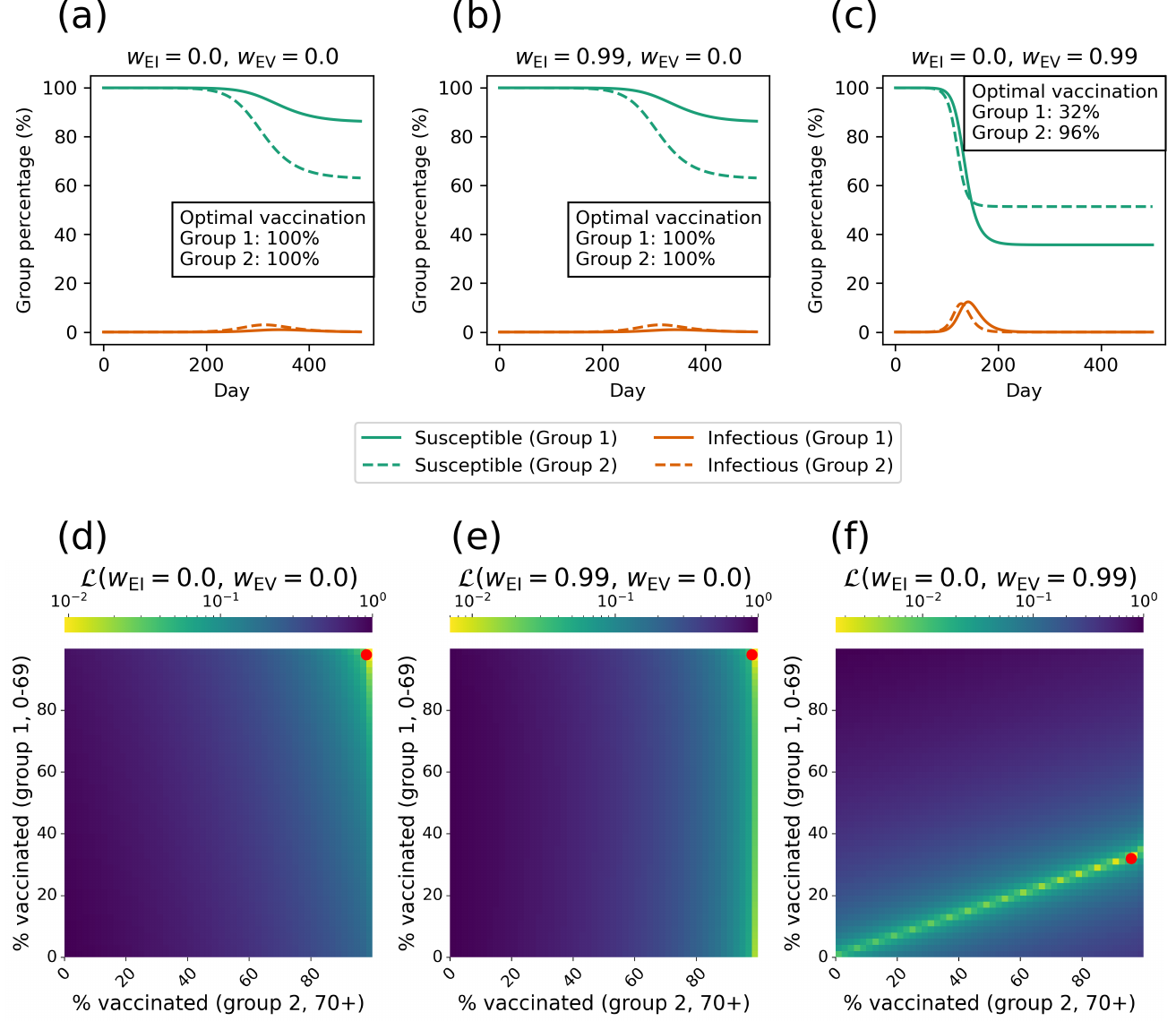}
\internallinenumbers
\caption{Examples of optimal vaccination strategies when vaccine supply is unlimited. Three different ethical frameworks are shown, one with emphasis on avoiding clinical burden (a, d), one placing priority on equity of infection-induced burden (b, e), and one with emphasis on equity of burden from adverse vaccination impacts (c, f). Trajectories generated under the optimal vaccination strategy for each framework are shown in (a), (b), and (c), while the corresponding loss surfaces are shown below, in (d), (e), and (f). In (d, e, f), red dots correspond to the optimal vaccination strategy. }
    \label{fig:traj_loss_unlimited}
\end{figure}

%Here we examine model results parameterised for COVID-19 as described above, and assume that the supply of vaccines is unlimited.
Epidemic trajectories under the optimal allocation of vaccines when there is an unlimited supply are shown in Figures \ref{fig:traj_loss_unlimited}a, \ref{fig:traj_loss_unlimited}b, and \ref{fig:traj_loss_unlimited}c for frameworks 1, 2 and 3, respectively.
To gain further insight into why these vaccination strategies are optimal under each of the ethical frameworks, Figures \ref{fig:traj_loss_unlimited}d, \ref{fig:traj_loss_unlimited}e and \ref{fig:traj_loss_unlimited}f show the global loss function (Equation \ref{eq:GLoss}) as a function of the vaccination proportions.
The optimal proportions (indicated with a red dot in Figures \ref{fig:traj_loss_unlimited}d, \ref{fig:traj_loss_unlimited}e and \ref{fig:traj_loss_unlimited}f) are the vaccination strategy used in the corresponding trajectory shown in Figures \ref{fig:traj_loss_unlimited}a, \ref{fig:traj_loss_unlimited}b, and \ref{fig:traj_loss_unlimited}c, respectively.

Full vaccination (100\% in groups 1 and 2) is selected as the optimal strategy for framework 1 and 2.
The trajectories in Figure~\ref{fig:traj_loss_unlimited}a and \ref{fig:traj_loss_unlimited}b demonstrate that, due to our parameterisation of vaccine efficacy and pathogen transmissibility, herd immunity is not achievable (substantial epidemics occur, even with 100\% vaccination).
% However, indirect protection through vaccination does affect the optimal strategy selection for Framework 2.
Comparing the loss surfaces for frameworks 1 and 2 (Figure~\ref{fig:traj_loss_unlimited}d, and \ref{fig:traj_loss_unlimited}e), both select full vaccination as the global optimum, however framework 2 (prioritising equity of disease burden) also has a local minimum at $(p_1 = 0,~p_2=1)$. 
This occurs because group 2 is at much higher risk of severe disease, and outcome equity can be achieved by withholding vaccines from group 1.
However, due to the indirect protection of group 2 achieved by vaccinating individuals in group 1, the full vaccination strategy is ultimately the best under both frameworks 1 and 2.
Thus, indirect protection avoids a potential value trade-off between equity in disease burden and aggregate clinical burden.

In contrast, because group 1 is at higher risk of adverse side-effects from vaccination, optimisation under framework 3 leads to vaccinating fewer individuals in group 1 and more infections in both groups 1 and 2 (Figure~\ref{fig:traj_loss_unlimited}c and \ref{fig:traj_loss_unlimited}f).
We note that the linear trend in Figure~\ref{fig:traj_loss_unlimited}f (a local maximum for $p_1<0.32$) follows the ratio established by the clinical burden due to adverse reactions: $C^V_1/C^V_2 = 1/3$.
This indicates a trade-off between equity in adverse impacts of vaccination, and aggregate clinical burden. 

\begin{figure}
    \centering
    \includegraphics[width=0.95\linewidth]{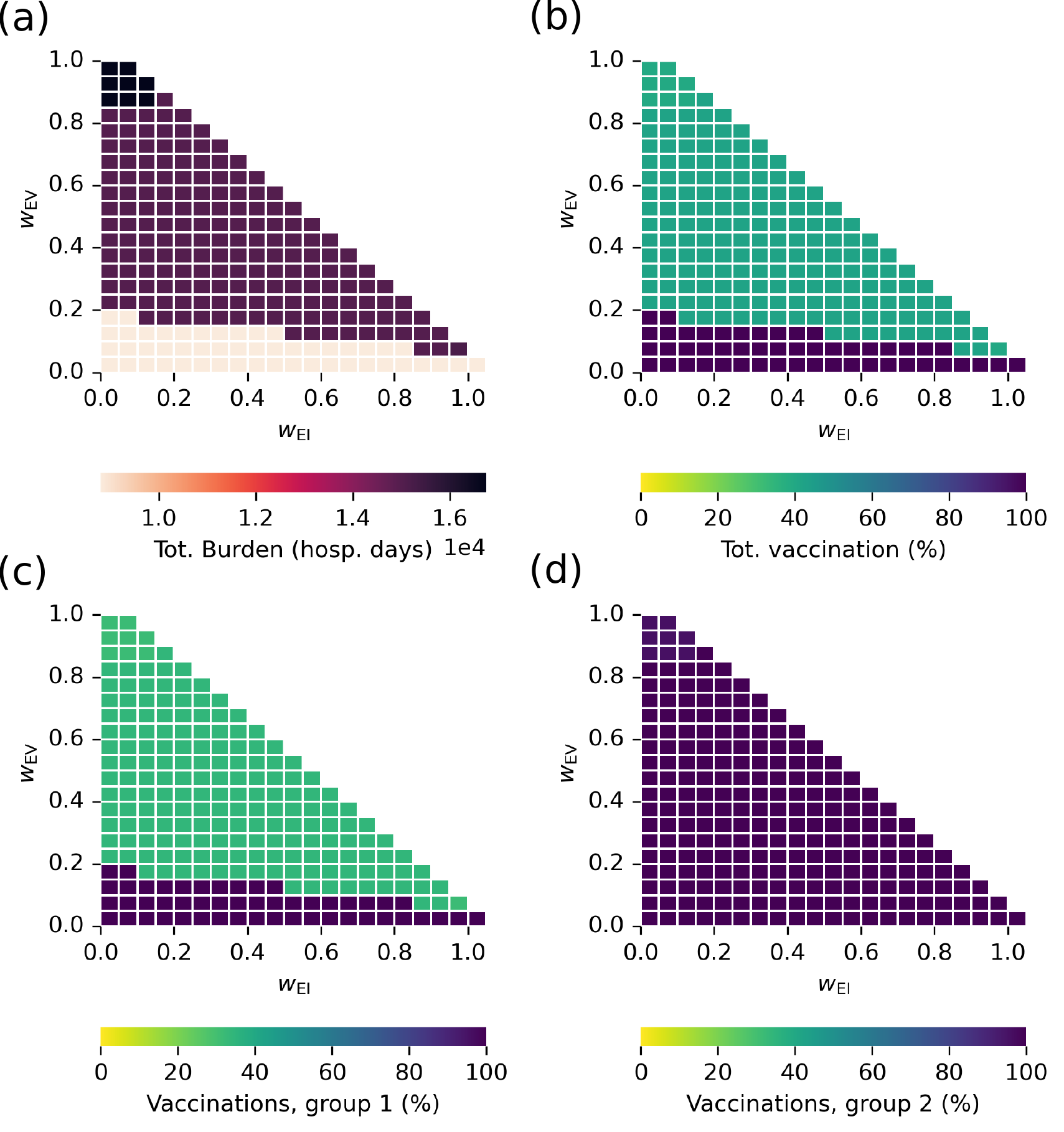}
    \internallinenumbers
    \caption{Outcomes of optimal vaccination strategies (with unlimited vaccine supply) as functions of the ethical frameworks used to define the loss function. Total clinical burden (days in hospital) is shown in (a). The proportion of the total population vaccinated is shown in (b). The proportion of individuals vaccinated in groups 1 and 2 are shown in (c) and (d), respectively.}
    \label{fig:hmaps_unlimited}
\end{figure}

Next, we examine how the optimal vaccination strategy differs over a broader range of value prioritisation. The results of optimal interventions with unlimited vaccine supply over intermediate values of $w_{\text{EI}}$ and $w_{\text{EV}}$ are shown in Figure~\ref{fig:hmaps_unlimited}.
The heatmap in Figure~\ref{fig:hmaps_unlimited}a shows the total clinical burden obtained when using the optimal vaccination for those values of $w_{\text{EI}}$ and $w_{\text{EV}}$.
The corresponding proportion of the total population vaccinated is shown in Figure\ref{fig:hmaps_unlimited}b. The allocation of vaccines is shown in Figure~\ref{fig:hmaps_unlimited}c and Figure~\ref{fig:hmaps_unlimited}d, for group 1 and group 2, respectively. 

Examining how the optimal vaccination strategy changes as a function of $w_{\text{EI}}$ and $w_{\text{EV}}$ allows us to understand how sensitive our allocation decision is to different prioritisation of values. Over almost all of the parameter space in $w_{\text{EI}}$ and $w_{\text{EV}}$, there are only two vaccine allocation strategies that are identified as optimal. These are shown in Figures \ref{fig:hmaps_unlimited}c and \ref{fig:hmaps_unlimited}d which provide optimal choice of vaccine allocation strategy for groups 1 and 2, respectively, as functions of $w_\text{EI}$ and $w_{\text{EV}}$. 

Given our chosen parameterisation based on SARS-CoV-2 and COVID-19, the results shown in Figure~\ref{fig:hmaps_unlimited}c and Figure~\ref{fig:hmaps_unlimited}d suggest the optimal vaccination strategy is insensitive to the priority given to equity in infection burden ($w_{\text{EI}}$).
However there is a clear decision boundary in Figure~\ref{fig:hmaps_unlimited}c as the weight put on equity in vaccine adverse effects ($w_{\text{EV}}$) increases.
At low levels, the optimal strategy is to vaccinate everyone in both groups 1 and 2, however, as more weight is placed on $\loss_{\text{EV}}$, there is a transition to vaccinate approximately 30\% of Group 1, leading to a higher clinical burden (Figure~\ref{fig:hmaps_unlimited}a). 

\FloatBarrier 
\subsection{Limited vaccine supply}

\begin{figure}
    \centering
    \includegraphics[width=0.95\linewidth]{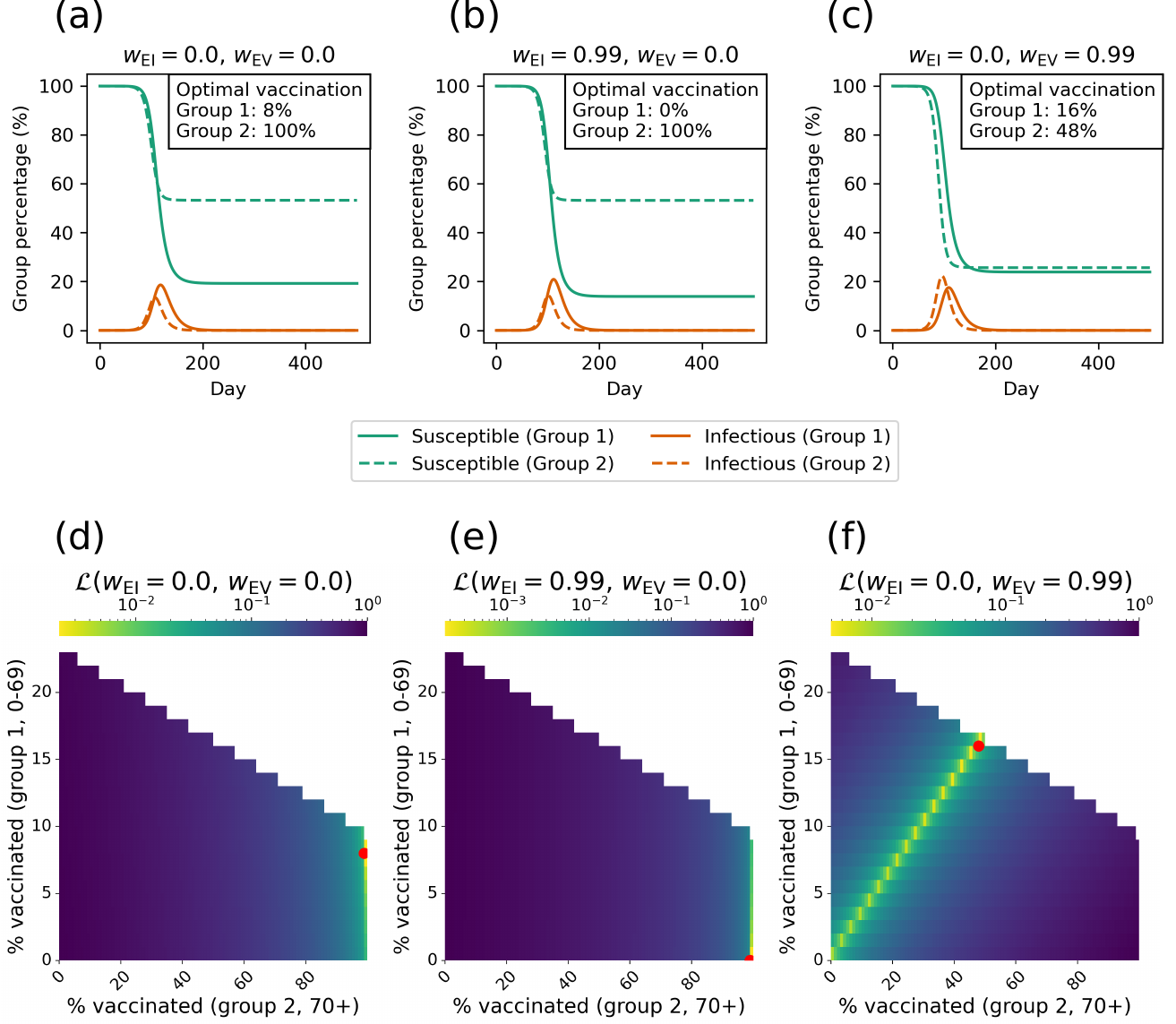}
    \internallinenumbers
    \caption{Examples of optimal vaccination strategies when vaccine supply is limited to $20\%$ of the population. Three different ethical frameworks are shown, one with emphasis on avoiding clinical burden (a, d), one placing priority on equity of infection-induced burden (b, e), and one with emphasis on equity of burden from adverse vaccination impacts (c, f). Trajectories generated under the optimal vaccination strategy for each framework are shown in (a), (b), and (c), while the corresponding loss surfaces are shown below, in (d), (e), and (f). In (d, e, f), red dots correspond to the optimal vaccination strategy. White space in (d, e, f) corresponds to strategies that exceed the vaccine supply.}
    \label{fig:traj_loss_limited}
\end{figure}

\begin{figure}
    \centering
    \includegraphics[width=0.95\linewidth]{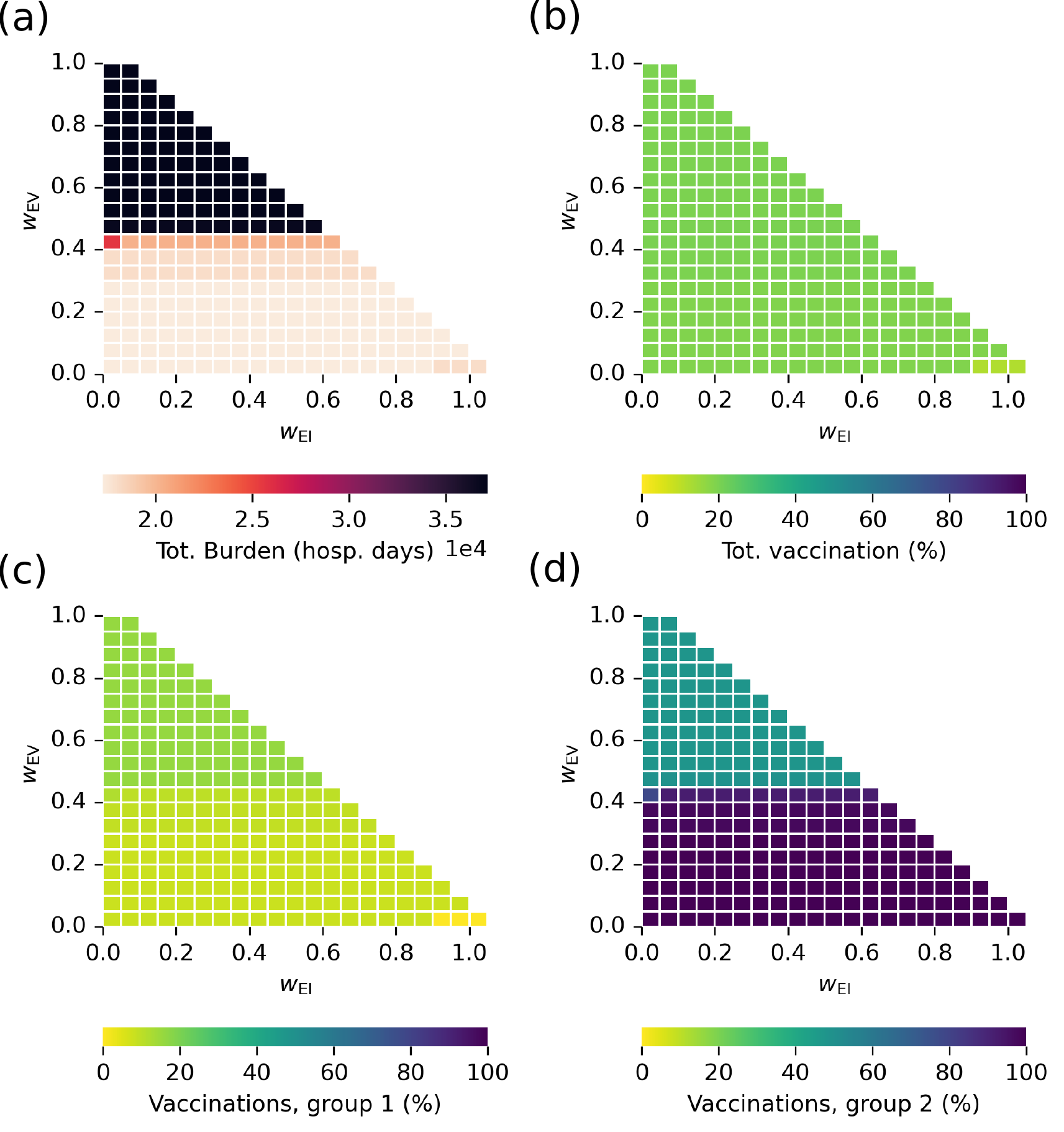}
    \internallinenumbers
    \caption{Outcomes of optimal vaccination strategies (with vaccine supply limited to $10^6$) as functions of the ethical frameworks used to define the loss function. Total clinical burden (days in hospital) is shown in (a). The total number of vaccines distributed is shown in (b). The proportion of individuals vaccinated in groups 1 and 2 are shown in (c) and (d), respectively.}
    \label{fig:hmaps_limited}
\end{figure}

Here we examine a scenario in which the supply of vaccines is limited to a maximum of $10^6$ (20\% of our total population).
As above, Figure~\ref{fig:traj_loss_limited} shows epidemic trajectories for frameworks 1--3 along with image plots of the corresponding loss surfaces. Heatmaps of outcomes of optimal strategies as a function of $w_{\text{EI}}$ and $w_{\text{EV}}$ are shown in Figure~\ref{fig:hmaps_limited}.

With limited vaccine supply, a tradeoff appears between equity of infection-associated clinical burden and aggregate clinical burden. While frameworks 1 and 2 both recommend vaccinating all individuals in group 2, the optimal strategies for group 1 are different: framework 1 recommends vaccinating the maximum number of individuals in group 1, given the number of vaccines remaining after vaccination of group 2 (Figure\ref{fig:traj_loss_limited}a,d). Alternately, framework 2 recommends no vaccinations for group 1, withholding available resources (Figure~\ref{fig:traj_loss_limited}b,e). On the other hand, framework 3 does not vaccinate all individuals in group 2, and instead spreads the vaccines between both groups (Figure~\ref{fig:traj_loss_limited}c,f). 

The choice of different strategies under frameworks 1 and 2 is different to the result observed for unlimited vaccine supply case, where the optimal strategy was to vaccinate everyone.
Unlike the scenario illustrated in Figures~\ref{fig:traj_loss_unlimited}b and \ref{fig:traj_loss_unlimited}e, here there are insufficient vaccines available to introduce indirect protection of group 2 by vaccinating group 1.
As such, the global maximum in Figure~\ref{fig:traj_loss_unlimited}e for the case of unlimited vaccine supply under framework 2 is not available and the optimal strategy is to withhold vaccines from group 1 to achieve a higher degree of similarity in the per-capita disease burden between groups.
This represents a trade-off between equity of disease outcomes and aggregate clinical burden.

Examining these tradeoffs over a broader range of ethical frameworks (Figure~\ref{fig:hmaps_limited}) reveals two distinct decision boundaries between three vaccination strategies. The trad-eoff between equity of infection-associated burden and aggregate burden demonstrated by framework 2 (compare Figure~\ref{fig:traj_loss_limited}d and \ref{fig:traj_loss_limited}e) is confined to extreme values of $w_{\text{EI}}$ (bottom right corner of the triangles in Figure~\ref{fig:hmaps_limited}). We note that this trade-off could be considered a violation of the principle of nonmaleficence, in that the choice is made to withhold available resources in order to promote equity of outcomes between two groups (by allowing the more well-off group to be subject to preventable burdens). However, this scenario appears only in extreme cases. For values of $w_{\text{EI}} \leq 0.8$, no tradeoff exists between aggregate clinical burden and equity of infection burden (there are no decision boundaries across the x-axis of the plots in Figure~\ref{fig:hmaps_limited}). On the other hand, a clear decision boundary exists as a function of $w_{\text{EV}}$. For values of $w_{\text{EV}} \geq 0.5$, higher aggregate burden is observed (Figure~\ref{fig:hmaps_limited}a) because fewer of the vulnerable individuals in group 2 are vaccinated (Figure~\ref{fig:hmaps_limited}d). Here, the vaccines have been distributed such that the burden of adverse events is distributed equitably between the two groups, despite the fact that this decision leaves many vulnerable individuals in group 2 without immunisation. This scenario is analogous to prioritising the equitable distribution of resources (this situation arises precisely if the adverse event rate per capita is identical in the two groups, $C^V_1=C^V_2$). When two groups have different risk profiles, equitable distribution of vaccines leads to inequitable outcomes. 
  
\FloatBarrier

\section{Discussion}

%\subsection{Statement of principal findings}

%\aez{I have edited to go for ``loss'' rather than ``objective''.}

We have presented an approach to defining loss functions for models of infectious disease interventions that include ethical values related to public health. These loss functions correspond to ethical frameworks, which, in the context of our study, we define as prioritisations of ethical values including reduction of aggregate clinical burden, and promotion of equity in clinical burdens between groups. By examining the strategies selected under different frameworks, we investigate potential value trade-offs. We constructed a simple model of infectious disease transmission and intervention (in this case, vaccination). We then defined a loss function to measure the aggregate clinical burden and inequity resulting from the intervention. We used the loss function to systematically locate the optimal intervention (i.e.~vaccination strategy) for different prioritisations of values. 

%\aez{I think to minimise the risk of misunderstanding/misrepresentation, we need to start with a strong statement that the model suggests vaccinate everyone you can (prioritising group 2) for all plausible regions of framework space. I've had a crack at this, but it comes out a little awkward in order to gel with the subsequent paragraph.}

There is a growing awareness that selecting interventions based on projections of aggregate clinical burden can produce inequitable outcomes or exacerbate existing inequities (see, e.g., \cite{tizzoni2022addressing}).
In our example based on COVID-19, we found for ethical frameworks that put a substantial weight on minimising aggregate burden, the optimal strategy was to vaccinate everyone, starting with those at most risk of severe disease. This was the optimal strategy even when some weight was given to outcome equity. We only found a trade-off between minimising aggregate clinical burden and outcome equity when vaccine supply was limited and the ethical framework placed negligible weight on aggregate burden.

%\aez{This statement about Stafford et al's work, does that mean, once you have enough for 30\%, then you are effectively in the unlimited vaccine scenario? I think the wording could be simplified to make this more explicit. I made some some edits to the subsequent sentences to streamline this, but it should be read by the person that introduced that reference.}

Our findings agree qualitatively with those of Stafford {\emph{et al.}}, who found (in a model of COVID-19 in the United States) that aggregate mortality and outcome equity could be jointly optimised as long as total vaccine coverage exceeded 30\% \cite{stafford2023retrospective}. 
While this outcome may be robust for scenarios describing vaccination against COVID-19, we expect it to be context specific. For example, a robust trade-off between equity of clinical burden and total clinical burden might be expected in scenarios where transmission is driven by subpopulations that are not at high risk of severe disease. In such cases, preventing low-risk infections may produce a high overall reduction in case numbers (through indirect protection), but could lead to an inequitable distribution of burdens because the group with high outcome risk is not directly protected. Such an alternate outcome is consistent with the findings of Goldstein {\it et al.} who investigated vaccine allocation strategies using a model of H1N1 influenza in the United States. They found that optimal strategies supported vaccination of children, who are at low risk of adverse outcomes but are disproportionately contagious if infected \cite{goldstein2010distribution}.

%\aez{This paragraph seemed a bit risky in terms of the potential for misunderstanding/representation so I have attempted to soften it and provide some additional context to the numbers we quote.}

On the other hand, we consistently observed a potential trade-off between aggregate clinical burden and the equitable distribution of intervention-induced harms. This trade-off was seen for both unlimited and limited vaccine supply as \(w_{\text{EV}}\) crossed a threshold at approximately \(0.1\) in the former and \(0.5\) in the latter. This can be considered a generalisation of the well-known trade-off between aggregate benefit and the uniformity of resource allocation in populations with heterogeneous risk profiles \cite{savulescu2020utilitarianism}. To put these observations into context, setting \(w_{\text{EV}}=0.5\) places at least as much weight on equity in adverse effects as is placed on reducing aggregate clinical burden, which may be unlikely to arise in practise. 

Another key observation was of abrupt decision boundaries in the space of value prioritisation (observable in Figures~\ref{fig:hmaps_unlimited} and \ref{fig:hmaps_limited}). Optimal strategies tended to correspond to one of the three extreme ethical frameworks: Frameworks 1--3.
This abruptness in the decision landscape is not guaranteed, and depends on the gradients of the loss surfaces for each value considered [see Figures \ref{fig:traj_loss_unlimited}(d,e,f) and \ref{fig:traj_loss_limited}(d,e,f)]. One interpretation of this feature is that, for the case we examined, ethical frameworks with nuanced balances of different values are not easily distinguished from extreme frameworks that mainly prioritise individual values. Again, we do not expect this to be a general feature of such decision spaces, and investigation of this property for other case studies is an avenue for future investigation.

\subsection{Limitations}

The simplicity of our transmission model implementation introduces several limitations. For example, a fixed parameterization of a deterministic model cannot address problems of uncertainty in model parameters or stochastic effects in the transmission process. Further, our assumption that vaccines are distributed prior to the initial outbreak means we could not investigate how the logistics of distribution may change the outcomes of a given strategy.

Our assessment of equity was also limited, because we elected to stratify the population into only two groups (those under the age of seventy, and those seventy or older). 
Age was chosen because it is clinically relevant in our example of COVID-19, but other important characteristics could be used as well, such as socio-economic status or ethnicity \cite{stafford2023retrospective}.
While quantifying equity over more complex population strata and multiple outcomes is a technical challenge, recent methods have been developed to achieve this through adaptation of Gini coefficients to stratified epidemic outcomes \cite{nguyen2023measuring}. While our study is constrained to only three values and a single parameterisation of a simple epidemic model, our approach is model-agnostic. 
Any system that can quantify the outcomes of prospective policy decisions could be compatible with loss functions like those we demonstrate here. The principal contribution of this study is to establish a link between frameworks of public health ethics, and the types of models used to inform intervention policy. 

We anticipate future methodological challenges when incorporating more comprehensive systems of values into increasingly complex models. To establish a proof of principle, we have treated both of these aspects with relative simplicity, which results in a number of limitations.
These are the other value considerations that were not included in our loss function. For example, inequity in access to resources; burdens arising from social interventions, such as stay-at-home orders; and unintended effects such as delayed medical treatments \cite{czeisler2020delay}.
The inclusion of state-of-the-art multi-criteria decision making approaches (e.g. \cite{harrison2024multi}) would allow a more realistic specification of values and constraints, and an enhanced capacity to capture nuanced aspects of ethical frameworks. 

While many ethical values in public health may potentially be included within modelling approaches such as this, there are some values that may not be quantifiable. For example, our loss function ignores preferences of the individuals in the population, it only considers what level of vaccination would be optimal based on quantifiable clinical burdens. As such, consideration of psychosocial harms introduced by the intervention are not included in our loss function. Relatedly, it cannot include any consideration of whether or not coercion would be justified to achieve intervention targets.

\section{Conclusion}
Frameworks of public health and biomedical ethics often seek to associate weights to principles through the process of value balancing \cite{beauchamp1994principles}. Our approach quantifies the effects of such weights in the context of a decision process made explicit by a mathematical model and optimisation problem. This facilitates exploration over the space of alternative weightings, to identify the presence, absence, and magnitudes of potential value trade-offs.
While it is beyond the scope of our method to offer specific guidance for selecting the value weights, our study provides insight into how robust the decision landscape is to changes in these weights. This work provides an approach to understanding how sensitive an optimal intervention strategy may be to a chosen set of value weights. By demonstrating value balancing in the context of a model with policy-relevant features, we hope to stimulate further work linking ethical frameworks to policy-relevant models of infectious disease interventions. 

\section{Data availability}
All code and data associated with this manuscript can be found on the GitHub repository:\\
\url{https://github.com/nefeltellioglu/ethical_framework}.

\section{Acknowledgements}
We thank the mathematical research institute MATRIX in Australia where part of this research was performed. Wasiur R. KhudaBukhsh acknowledges funding from the MATRIX-Simons Travel Grant and the Engineering and Physical Sciences Research Council (EPSRC) [grant number EP/Y027795/1]. Jessica E. Stockdale acknowledges funding from the MATRIX-Simons Travel Grant and the Natural Science and Engineering Research Council (Canada) [Discovery Grant RGPIN-2023-04509]. Julie A. Spencer acknowledges funding from the Laboratory Directed Research and Development program of Los Alamos National Laboratory under project number 20240066DR.

\bibliographystyle{unsrt}  
%\section*{References}
\bibliography{references} 

\FloatBarrier
\clearpage
\newpage 

\newcommand{\beginsupplement}{%

 \setcounter{table}{0}
   \renewcommand{\thetable}{S\arabic{table}}%
   
     \setcounter{figure}{0}
      \renewcommand{\thefigure}{S\arabic{figure}}%
      
      \setcounter{page}{1}
      \renewcommand{\thepage}{S\arabic{page}} 
      
      \setcounter{section}{0}
      \renewcommand{\thesection}{S\arabic{section}}
      
      \setcounter{equation}{0}
      \renewcommand{\theequation}{S\arabic{equation}}
     }

\beginsupplement

\section{Supplementary Information}
\vspace{1.0cm}

\subsection{Supplemental methods}
%CZ: I'm moving this to the supplement: 
\subsubsection{The effective reproduction number}

Using the next-generation matrix \cite{diekmann2010construction} method, we can get the basic reproduction number of our model as
\begin{equation*}
    \Ro = \frac{\beta _{11}+\beta _{22} + \sqrt{\beta _{11}^2-2 \beta _{22} \beta _{11}+\beta _{22}^2+4 \beta _{12} \beta _{21}}}{2 \gamma }\,.
\end{equation*}

In our implementation, we specify $\Ro$,  based on estimates for the effective reproduction number of the Omicron variant of COVID-19, and compute a global transmission scalar $\beta$ as: 
\begin{equation}
\beta = \Ro \gamma \bigg[ c_{11} + c_{22} + \sqrt{c_{11}^2 - 2 c_{22} c_{11} + c_{22}^2 + 4 c_{12} c_{22}} ~\bigg]^{-1}
\end{equation}
such that 
\begin{equation}
\beta_{ij} = \beta c_{ij}
\end{equation}

\subsubsection{ODE model (SIR with imperfect vaccination)}

\begin{equation}\label{eq:ode_model_with_vacc}
  \begin{split}
  \frac{dS_1}{dt} &=  - \left( \beta_{11} (I_1+I_{1,U}) + \beta_{21} (I_2+I_{2,U}) \right)\frac{S_1}{N_1}\\
  \frac{dS_{1,U}}{dt}&= - \left( \beta_{11} (I_1+I_{1,U}) + \beta_{21} (I_2+I_{2,U}) \right)\frac{S_{1, U}}{N_1}\\
  \frac{dS_{1,P}}{dt} &= 0\\
  \frac{dI_1}{dt}&=  \left(\beta_{11} (I_1+I_{1,U}) + \beta_{21} (I_2+I_{2,U})\right)\frac{S_1}{N_1} - \gamma I_1\\
  \frac{dI_{1,U}}{dt}&=  \left(\beta_{11} (I_1+I_{1,U}) + \beta_{21} (I_2+I_{2,U})\right)\frac{S_{1,U}}{N_1} - \gamma I_{1,U}\\
  \frac{dR_1}{dt}&= \gamma I_1 \\
  \frac{dR_{1, U}}{dt}&= \gamma I_{1,U} \\
  \frac{dS_2}{dt}&=- \left(\beta_{22} (I_2+I_{2,U}) + \beta_{12} (I_1+I_{1,U}) \right)\frac{S_2}{N_2} \\
  \frac{dS_{2,U}}{dt}&=- \left(\beta_{22} (I_2+I_{2,U}) + \beta_{12} (I_1+I_{1,U}) \right)\frac{S_{2,U}}{N_2} \\
  \frac{dS_{2,P}}{dt}&= 0\\
  \frac{dI_2}{dt}&= \left(\beta_{22} (I_2+I_{2,U}) + \beta_{12} (I_1+I_{1,U}) 
\right)\frac{S_2}{N_2} - \gamma I_2 \\
\frac{dI_{2,U}}{dt}&= \left(\beta_{22} (I_2+I_{2,U}) + \beta_{12} (I_1+I_{1,U}) 
\right)\frac{S_{2,U}}{N_2} - \gamma I_{2,U} \\
  \frac{dR_2}{dt}&= \gamma I_2 \\
  \frac{dR_{2,U}}{dt}&= \gamma I_{2,U} \\
   N_1 &= S_1+I_1+R_1 + S_{1,U}+I_{1,U}+R_{1,U}+S_{1,P}\\
   N_2 &= S_2+I_2+R_2 + S_{2,U}+I_{2,U}+R_{2,U}+S_{2,P}
  \end{split}  
\end{equation}

\subsection{Parameterisation for COVID-19}

Below are detailed descriptions of how the parameters in Table \ref{tab:sir2-params} were derived: 

\begin{itemize}
\item{$N_1$, $N_2$ subpopulation sizes corresponding to age groups 0--69 and 70+ were taken from the Australian Bureau of Statistics demographic data (Table\ref{tab:params1}).}

\item{$\Ro$ the basic reproductive ratio corresponds to the effective reproduction number of the Omicron variant of SARS-CoV-2 (Table \ref{tab:params1}). }

\item{contact rates between subpopulations $c_{ij}$ were derived using the R package socialmixr from the POLYMOD contact survey of the United Kingdom, aggregated to the age groups [0, 69] and 70+ (Table \ref{tab:params1}).}

\item{$\gamma$ the infection recovery rate was computed as the sum of the mean incubation period and mean symptom recovery period estimated for the Omicron variant of SARS-CoV-2 (Table \ref{tab:params1}).}

\item{$\text{VE}_I$ vaccine efficacy against infection was takend directly from a literature estimate (see Table \ref{tab:params2}).}

\item{$\text{VE}_{SD|I}$ vaccine efficacy against severe disease given infection was computed from independent estimates of $\text{VE}_I$ and $\text{VE}_{SD}$ (see Table \ref{tab:params2}).}

\item{$C^I_1$ the average clinical burden (days in hospital) for an infection in group 1 (aged 0--69) was computed as the population-weighted sum of the infection hospitalisation rate (IHR) over age groups between 0 and 69 years (see Table \ref{tab:params3}) multiplied by the population-weighted sum of the mean age-stratified length of stay (LoS) in hospital due to infection (Note different age strata were used for the IHR and LoS components, see Table \ref{tab:params3}).}

\item{$C^I_2$, the average clinical burden (days in hospital) caused by an infection in group 2 (aged 70+) was computed using the same approach as $C^I_1$, but using age strata in the 70+ range (Table \ref{tab:params3}).}

\item{$C_1^V$ the average clinical burden caused by vaccination of a member of group 1 (aged 0--69) was computed as the population-weighted mean of adverse event rates for men under 30 years of age (approx. 10 times higher than that of the general population), and event rates for the rest of the population under 69 years old, multiplied by the mean estimate of hospital LoS from vaccine-associated myocarditis (Table \ref{tab:params2}).}

\item{$C_2^V$ was calculated as the rate of vaccine-associated myocarditis in the general population (not including men aged under 30 years), multiplied by the mean hospital LoS for vaccine-associated myocarditis (Table \ref{tab:params2}). }

\end{itemize}

\begin{table}[ht]
    \centering
    \begin{tabular}{m{5cm}|m{4cm}|m{4cm}}
    \hline
       Model component - \newline \textbf{Basic transmission model} & Value & Source \\
    \hline

      % CZ verified
      Population size & 5 million & \makecell[l]{Melbourne population \\ (ABS 2024, rounded \\ to nearest million \cite{MelPop_ABS2024})}\\

      % CZ verified
      Age groups [\% total pop.] &  \makecell[l]{0-69 years [87.9\%], \\  70+ years [12.1\%]} & Australia, June 2023 \cite{AusPop_ABS2024}  \\

      %CZ verified
       $\Ro$ (to derive $\beta$)  & \makecell[l]{Basic: 9.5 \\ Effective: 3.4} & Omicron variant \cite{liu2022effective} \\

        % CZ calculated
       \makecell[l]{$c_{ij}$ group-wise \\ contact multipliers} &  $\begin{bmatrix} 0.38 & 0.14 \\  0.14 &0.34 \end{bmatrix} $   &  POLYMOD data from the UK processed in R using socialmixr \cite{funk2024introduction,mossong2008social}\\

       % incubation period 
       Incubation period & 3.5 days & Taiwan (Omicron BA.5, 2022) \cite{cheng2024sars}\\

       % symptom recovery period 
       Symptom recovery period &  6.87 days & United Kingdom (Omicron, 2021-2022) \cite{menni2022symptom} \\

       % I recommend we add the incubation period to this. 
       Recovery period ($\gamma^{-1}$)& \makecell[l]{ $6.87 + 3.5 = 10.37$ days} & - \\
       
       \hline
\end{tabular}
\internallinenumbers
    \caption{Parameters for the transmission model, focused on the COVID-19 Omicron variant, July 2021 - Feb 2022. }
    \label{tab:params1}
\end{table}

\begin{table}[ht]
    \centering
    \begin{tabular}{m{5cm}|m{4cm}|m{4cm}}
    \hline
       Model component - \newline \textbf{Vaccination}  & Value & Source \\
    \hline
        
        Efficacy against infection ($\text{VE}_{I}$) & 0.531 (Omicron)  & \makecell[l]{Systematic review \\ (Song {\textit{et al.}} \cite {song2023effectiveness})}\\

        Efficacy against severe disease ($\text{VE}_{SD}$) & 0.825 (Omicron) & \makecell[l]{Systematic review \\ (Song {\textit{et al.}} \cite {song2023effectiveness})}\\
        
        \makecell[l]{\\ Efficacy against severe disease \\ given infection ($\text{VE}_{SD|I}$)} & $1 - \frac{1 - \text{VE}_{SD}}{1 - \text{VE}_{I}} = 0.627$ & -\\
        
        \makecell[l]{Rate of severe adverse effects \\ (baseline)} & 2/$10^5$ events per dose & Myocarditis in general population (vaccine safety report, Australian TGA, 2022 \cite{vaccinesafety2022}) \\

        \makecell[l]{Rate of severe adverse effects \\ (men aged under 30)} & 20/$10^5$ events per dose & Myocarditis in men aged under 30 (vaccine safety report, Australian TGA, 2022) \cite{vaccinesafety2022}\\
        
        \makecell[l]{Cost of severe effects \\ (mean days hospitalised)} & 5.7 days & Vaccine-associated myocarditis (BNT162b2) \cite{patone2022risks} \\
         \hline
    \end{tabular}
    \internallinenumbers
    \caption{Resources to inform parameters for the vaccination model, focused on Omicron COVID-19 waves through July 2022.}
    \label{tab:params2}
\end{table}

\begin{table}[ht]
    \centering
    \begin{tabular}{m{5cm}|m{4cm}|m{4cm}}
    \hline
       Model component - \newline \textbf{Clinical burden} & Value & Source \\
    \hline

    Average hospital stay (non-ICU) days  & \makecell[l]{$[0,39]: 2.16$\\ $[40,69]:3.93$\\ $70+:7.61$} & New South Wales (Omicron) \cite{tobin2023real} \\

       %case hospitalisation rate & 0.92\% & Washington State \cite{paredes2022associations} \\

        Infection hospitalisation rate by age & \makecell[l]{
        Total: 0.22\% \\  
        $[0,4]:$ 0.41\% \\
        $[5,9]:$ 0.10\% \\
        $[10,19]:$ 0.013\% \\
        $[20,29]:$ 0.022\% \\
        $[30,39]:$ 0.04\% \\
        $[40,49]:$ 0.03\% \\
        $[50,59]:$ 0.06\% \\
        $[60,69]:$ 0.25\% \\
        $[70,79]:$ 0.42\% \\
        $[80+]:$ 2.17\%} & 
        British Columbia, Omicron (BA.2, BA.5) \cite{skowronski2023risk} \\ 
     \hline
    \end{tabular}
    \internallinenumbers
    \caption{Resources to inform parameters for the burden model, focused on Omicron COVID-19 waves through July 2022. }
    \label{tab:params3}
\end{table}

\end{document}